# BLACK HOLE CENSORSHIP OF VARYING FUNDAMENTAL CONSTANTS

JANE H. MACGIBBON
Dept of Physics and Chemistry
University of North Florida
Jacksonville FL 32224

February 2006

ABSTRACT

**Here it is shown that the variation in the fine structure constant measured by Webb et al. matches the theoretically derived value for the maximum variation in the electronic charge permitted by the Generalized Second Law of Thermodynamics for black holes accreting and emitting in the present cosmic microwave background. It is postulated that the constants of nature, independently or dependently, vary at the maximal rate allowed by the Generalized Second Law of Thermodynamics.**

Observations[1,2] of absorption in the spectra from distant quasars intriguingly suggest that the fine-structure constant $\alpha$, which governs electromagnetic interactions, may be increasing as the Universe ages. The observations are consistent with a rate of change of roughly $\Delta\alpha/\alpha \sim 2 \times 10^{-23}$ per second. The fine structure constant $\alpha = e^2/\hbar c$ where $e$ is the charge of the electron, $\hbar$ is Planck's constant and $c$ is the speed of light. In a recent Brief Communications P.C.W. Davies et al. have argued[3], using black hole thermodynamics, that an increase in $\alpha$ can not be due solely to an increase in the electronic charge $e$. The authors reasoned that an increase in $e$ would simultaneously decrease the entropy of an electrically charged microscopic black hole which is emitting Hawking radiation. They found that this entropy decrease is not compensated by a greater increase in the entropy of the black hole environment. The authors claimed that such a net entropy decrease would violate the Second Law of Thermodynamics.

Here we note, however, that the Second Law of Thermodynamics requires that the net entropy of the system increases or is constant *over any time interval*. In contrast, Davies et al interpreted the entropy constraint to mean that the entropy of the system is not decreased when $e$ is increased *at a fixed time*. Thus they neglected the entropy change due to Hawking radiation which is occurring over any time interval.

In this paper, we include the full description of the time variation of the entropy of the black hole system. When fully described, we shall see that a small increase in $e$ of $\Delta e/e = \Delta\alpha/2\alpha \sim 10^{-23}$ per second does not violate the generalized entropy law for black holes in the present Universe. Thus black hole thermodynamical constraints do not rule out the possibility that an increase in $\alpha$ is due solely to an increase in electric charge $e$. Furthermore we will discover that $de/dt \approx 10^{-23} e$ per second matches the maximum variation in $e$ permitted for black holes in the present cosmic microwave background.

Throughout this paper we assume that $c$, $\hbar$ and the gravitational constant $G$ are constant. Models with dependent variation will be treated elsewhere.

\*     \*     \*

The Generalized Second Law of Thermodynamics, derived for black hole systems, states that the net entropy of the system can not decrease with time[4]. Over a time interval $\Delta t$, the net generalized entropy of the system increases by

$$\Delta S = \Delta S_{BH} + \Delta S_{R+M} \geq 0 \tag{1}$$

where $\Delta S_{BH}$ and $\Delta S_{R+M}$ are the change in entropy of the black hole and of the ambient radiation and matter, respectively. The entropy of a black hole is

$$S_{BH} = \frac{kc^3}{4\hbar G} A_{BH} \tag{2}$$

where $k$ is the Boltzmann constant. For a charged, non-rotating (Reissner-Nordstrøm) black hole of mass $M$ and charge $Q$ in esu units, the area of the black hole is

$$A_{BH} = \frac{4\pi G^2}{c^4}\left(M + \sqrt{M^2 - Q^2/G}\right)^2 \tag{3}$$

Hawking[5,6] has established that a black hole is continuously emitting quasi-thermal radiation with a temperature

$$T_{BH} = \frac{2\hbar G}{kc}\frac{\sqrt{M^2 - Q^2/G}}{A_{BH}} = \frac{\hbar c^3}{2\pi kG}\frac{\sqrt{M^2 - Q^2/G}}{\left(M + \sqrt{M^2 - Q^2/G}\right)^2}. \tag{4}$$

Thus $\Delta S_{BH}$, the full change in black hole entropy over time $\Delta t$, must include the contribution from the Hawking flux as well as any partial change induced by a variation in the electronic charge, i.e.

$$\Delta S_{BH} \approx \frac{dS_{BH}}{dt}\Delta t = \frac{kc^3}{4\hbar G}\left(\frac{\partial A_{BH}}{\partial t} + \frac{\partial A_{BH}}{\partial e}\frac{de}{dt}\right)\Delta t \tag{5}$$

Only the second term was considered by Davies et al..

In the general case,

$$\frac{\partial A_{BH}}{\partial t} = \frac{8\pi G^2}{c^4}\frac{\left(M + \sqrt{M^2 - Q^2/G}\right)}{\sqrt{M^2 - Q^2/G}}\left\{\left(M + \sqrt{M^2 - Q^2/G}\right)\frac{dM_H}{dt} - \frac{Q}{G}\frac{\partial Q_H}{\partial t}\right\} \tag{6a}$$

$$\frac{\partial A_{BH}}{\partial e} = \frac{8\pi G^2}{c^4}\frac{\left(M + \sqrt{M^2 - Q^2/G}\right)}{\sqrt{M^2 - Q^2/G}}\left\{-\frac{Q}{G}\frac{\partial Q}{\partial e}\right\} \tag{6b}$$

where $\partial Q/\partial e = Q/e$ (if $\partial Q/Q = \partial e/e$) and the subscript $H$ denotes Hawking radiation. Both $M$ and $Q$ change as the black hole radiates. To proceed further we must consider the two cases when the black hole temperature is greater than and less than the temperature of its surroundings.

Case (I) If the black hole temperature is greater than the temperature of its surroundings, ie $T_{BH} > T_{R+M}$, there will be a net radiation loss from the black hole into its environment.

Case (IA) Consider first the case when $Q$ is not affected by the Hawking radiation, i.e. the black hole temperature is below about 100 keV, the threshold to emit the lightest charged particle, the electron.

This corresponds to $M \gtrsim 10^{17}$ gm. Then $\partial Q_H / \partial t = 0$ and so

$$\frac{\partial A_{BH}}{\partial t} = \frac{\partial A_{BH}}{\partial M}\frac{dM}{dt}$$

and

$$\frac{dS_{BH}}{dt} = \frac{2\pi k G}{\hbar c}\left[1+\sqrt{1-Q^2/GM^2}\right]^{-1}\left\{\left(M+\sqrt{M^2-Q^2/G}\right)\frac{dM_H}{dt} - \frac{Q}{G}\frac{\partial Q}{\partial e}\frac{de}{dt}\right\} \quad (7)$$

The mass loss due to Hawking radiation is[7]

$$\frac{dM_H}{dt} \approx -\frac{\hbar c^4}{G^2 M^2}\beta \quad (8)$$

with $\beta \approx 3\times 10^{-4}$ for a hole emitting the photon, 3 light neutrino species and the graviton. Strictly the mass loss rate (8) applies only if $Q \ll Q_{MAX} = G^{1/2}M$, the maximal possible charge on a black hole. This will suffice for our purpose because, as described below, high $Q$ is discharged quickly[8,10,11,12] provided that $M \lesssim \hbar ce/G^{3/2}m_e^2 \approx 10^5 M_\odot$ where $m_e$ is the electron mass. D.N. Page has numerically calculated[9] that for a black hole emitting the photon, 3 light neutrino species and the graviton the increase in $S_{R+M}$ due to Hawking emission is 1.62 times the corresponding decrease in $S_{BH}$ due to Hawking emission. Thus $\Delta S \geq 0$ provided the second term within the {} brackets in Eq (7) is not of order the first term. For $de/dt \approx 10^{-23} e$ per second, the second term is of order the first term when the black hole has a charge of

$$Q_{1\approx 2} \approx \left\{\frac{\hbar c^4 \beta}{GM\left(e^{-1}de/dt\right)}\right\}^{1/2}\left(2-\frac{\hbar c^4 \beta}{G^2 M^3\left(e^{-1}de/dt\right)}\right)^{1/2}$$

$$\approx 6\times 10^{20}\left(M/\text{gm}\right)^{-1/2}\left\{2-\left(\frac{M}{1.8\times 10^{16}\text{g}}\right)^{-3}\right\}^{1/2} \text{esu} \quad (9)$$

For $M \gtrsim 1.8\times 10^{16}$ gm, $Q_{1\approx 2} \lesssim Q_{MAX}$ and so $Q_{1\approx 2}$ may be achievable. However, Gibbons[10] and Zaumen[11] have shown that if the charge is greater than $Q_{PP} \approx G^2 m_e^2 M^2/\hbar ce$ where $m_e$ is the electron mass, the black hole will quickly discharge by superradiant[13,14] Schwinger-type $e^+e^-$ pair-production in the electrostatic field surrounding the hole. From Eq (6a) this Schwinger-type discharge will increase $\partial A_{BH}/\partial t$ and hence $S_{BH}$, as well as $S_{R+M}$. The condition $Q_{PP} \lesssim Q_{1\approx 2}$ is satisfied by all holes lighter than

$$M_{PP} \approx \left(\frac{2\hbar^3 c^6 e^2 \beta}{G^5 m_e^4\left(e^{-1}de/dt\right)}\right)^{1/5} \approx 3\times 10^{25} \text{ gm}. \quad (10)$$

Gibbons has derived the discharge rate for $T_{BH} \ll 100$ keV black holes with $Q > Q_{PP}$ to be[10]

$$\frac{dQ_{PP}}{dt} \approx \frac{Q^3 e^4}{\hbar^3 c^2 r_+^2} \exp\left(-\frac{\pi c^3 m_e^2 r_+^2}{\hbar Q e}\right) \tag{11}$$

where $r_+ = G\left(M + \sqrt{M^2 - Q^2/G}\right)/c^2$. The pair-production discharge term $(Q/G)dQ_{PP}/dt$ from a $Q_{1\approx 2}$ black hole is greater than the $(Q/G)(\partial Q/\partial e)de/dt$ term in Eq (7) for all $M \lesssim M_{PP\approx E}$ where $M_{PP\approx E}$ satisfies

$$\frac{c^4 e^4 \beta}{\hbar^2 G^2 M^2 \left(e^{-1} de/dt\right)^2} \approx \exp\left(\frac{4\pi G^{5/2} m_e^2 M^{5/2} \left(e^{-1} de/dt\right)^{1/2}}{\sqrt{2}\hbar^{3/2} c^3 e \beta^{1/2}}\right) \tag{12}$$

ie $M_{PP\approx E} \approx 7.0 \times 10^{25}$ gm

Because the pair production term grows much faster with $Q$ than the $de/dt$ term, the pair production term then dominates when $Q_{1\approx 2} \lesssim Q$ for $M \lesssim M_{PP\approx E}$. To the accuracy of our analysis and the original references (which together may be roughly $\pm$ a factor of 2) and the measurements of $\Delta\alpha/\alpha$, our estimate of $M_{PP,PP=E} \approx 3-7 \times 10^{25}$ gm coincides with $M_{CMB} \approx 4.5 \times 10^{25}$ gm, the mass of a black hole whose temperature is equal to the ambient temperature of the Universe $T = 2.73$ K. This coincidence is remarkable given that the possible mass range for black holes in the current Universe extends over at least 50 orders of magnitude, from the Planck mass $\sim 10^{-5}$ gm to at least the mass of an AGN core $\sim 10^{45}$ gm. It raises the question do $\alpha$ and $e$ vary at the maximal rate allowed by the Generalized Second Law of Thermodynamics? The Webb et al. quasar data show an increase in $\alpha$ of a few parts per million over the past 4 - 12 billion years when compared with the present value of $\alpha$. Over that timespan, $M_{CMB}$ has increased from $M_{CMB} \approx 1.1 \times 10^{25}$ gm at 12 billion years ago to $M_{CMB} \approx 3.6 \times 10^{25}$ gm at 4 billion years ago to $M_{CMB} \approx 4.5 \times 10^{25}$ gm today. Furthermore, the Webb et al. measurements suggest that the rate of increase in $\alpha$ is weakening as the Universe ages, again consistent with Eqs (10) and (12).

Thus in case (IA), the net increase in $S_{BH} + S_{R+M}$ due to Hawking emission and pair production is greater than the decrease in $S_{BH}$ induced by an electronic charge change of $de/dt \approx 10^{-23} e$ per second for all neutral and charged black holes whose temperature is greater than the 2.73K cosmic microwave background. However the entropy constraint would be violated today if the Universe were an order of magnitude cooler.

Case (IB) Consider the case when the black hole is emitting charged particles via Hawking emission, ie $\partial Q_H/\partial t \neq 0$ and $M \lesssim 10^{17}$ gm. A charged black hole emits its charge at a rate which depends on $Q$ and preferentially emits particles of the same sign as its own charge. Then

$$\frac{\partial Q_H}{\partial t} = -\frac{|Q|}{Q} e \frac{dN_H}{dt} \tag{13}$$

where $e dN_H/dt$ is the net emission rate of charge out of the black hole. If $E_{av} \approx 5 T_{BH}$ is the average energy[7] of a particle emitted by the black hole and $dN_{TOT}/dt$ is the total emission rate of all particles from the black hole then $dM_H/dt = -(E_{av}/c^2)dN_{TOT}/dt \leq 0$ and

$$\left| Q \frac{\partial Q_H}{\partial t} \right| \leq \frac{c^2 e}{E_{av}} \left| Q \frac{dM_H}{dt} \right| \qquad (14)$$

Thus $dS_{BH(IB)}/dt$ lies in the range

$$\frac{2\pi k G}{\hbar c}\left[1+\sqrt{1-Q^2/GM^2}\right]^{-1}\left\{\left(M+\sqrt{M^2-Q^2/G} - \frac{8\pi M|Q|e}{5\hbar}\right)\frac{dM_H}{dt} - \frac{Q}{G}\frac{\partial Q}{\partial e}\frac{de}{dt}\right\} \geq \frac{dS_{BH(IB)}}{dt}$$

$$\geq \frac{2\pi k G}{\hbar c}\left[1+\sqrt{1-Q^2/GM^2}\right]^{-1}\left\{\left(M+\sqrt{M^2-Q^2/G}\right)\frac{dM_H}{dt} - \frac{Q}{G}\frac{\partial Q}{\partial e}\frac{de}{dt}\right\} \qquad (15)$$

For $Q \ll Q_{MAX}$ it is straightforward to show that the net entropy increase due to the Hawking emission, given by Eq (8) with[15] $\beta \approx 4 \times 10^{-4}$ for $M \lesssim 10^{17}$ gm, dominates the entropy decrease due to $de/dt \approx 10^{-23} e$ per second. For higher $Q$, the relevant Hawking emission rate per degree of particle freedom of spin $s$ particles with energy in the range $(E, E+dE)$ is

$$d\dot{N} = \frac{\Gamma_s dE}{2\pi\hbar}\left\{\exp\left(\frac{E-c^2 eQ/GM}{kT_{BH}}\right) - (-1)^{2s}\right\}^{-1} \qquad (16)$$

where $\Gamma_s$ is the spin- and charge-dependent absorption probability. As $Q$ increases, the emission rate is modified by the electrostatic chemical potential term in Eq (16). Carter has estimated[12] that $dQ_e/dt \approx -c^2 e^2 Q/\hbar G M$ for $M \lesssim 10^{17}$ gm and $Q \leq \hbar c/e$ (the thermal regime) and $dQ_e/dt \approx -e^4 Q^3/\hbar^3 GM$ for $Q \geq \hbar c/e$ (the superradiant regime) which matches on to Eq (11) at higher $M$. In both cases the entropy increase due to the $dQ_e/dt$ term dominates the decrease due to $de/dt \approx 10^{-23} e$ per second for all $M \lesssim 10^{17}$ gm. We should also note that[12] the discharge timescale $\tau_Q \approx Q/\dot{Q}_e$ is much smaller than the black hole lifetime $\tau_{BH} \approx G^2 M^2/\hbar c^4$ for $M \lesssim 10^{17}$ gm and is even comparable with or less than $\hbar c/e^2 = 137$ times the characteristic timescale for a black hole to form $\tau_F \approx r_+/c$. This implies[8,10] that $M \lesssim 10^{17}$ gm black holes should be essential neutral today, up to random fluctuations of order the Planck charge $(\hbar c)^{1/2}$.

Thus for all neutral and charged black holes in case (IB), there is a net increase in $S_{BH} + S_{R+M}$ if $de/dt \approx 10^{-23} e$ per second.

Case (II) If the black hole temperature is less than or equal to the temperature of its surroundings, i.e. $T_{BH} \leq T_{R+M}$ (for example if the black hole is placed in a same temperature heat bath as suggested by Davis et al), the net entropy also increases when $e$ increases at the rate indicated by the Webb et al observations. This can be shown by explicitly deriving the heat flow into the hole etc and/or by general thermodynamical principles as follows.

An increase in $e$ will decrease the black hole area $A_{BH}$ and temperature $T_{BH}$. Once $T_{BH}$ drops below the ambient temperature, the black hole will accrete from its surroundings faster than it Hawking radiates. This accretion increases the black hole mass $M$, further lowering $T_{BH}$, and leads in turn to more accretion. (As Hawking has pointed out[4], a black hole can not be in stable thermal equilibrium if an unbounded amount of energy is available in its surroundings. This also means that the Davies et al suggestion that a black hole can be kept in isoentropic equilibrium with a same temperature heat bath is not achievable.) The general thermodynamical definition of the temperature of the environment is

$$T_{R+M}^{-1} \equiv \frac{\partial S_{R+M}}{\partial E} \qquad (17)$$

where $E$ is the energy, and the definition of the black hole temperature is[4]

$$T_{BH}^{-1} \equiv c^{-2}\left(\frac{\partial S_{BH}}{\partial M}\right)_{Q\,\text{fixed}} \qquad (18)$$

It can readily checked that the explicit form of $T_{BH}$ given in Eq (4) satisfies this definition.

During accretion, the black hole mass increases by an amount equal to the decrease in the energy of the environment. Hence for $T_{BH} \leq T_{R+M}$, the temperature definitions imply that the increase in black hole entropy due to accretion must be greater than the decrease in $S_{R+M}$ due to accretion. Also for $T_{BH} \leq T_{R+M}$, the increase in $S_{BH}$ due to accretion must be greater than the decrease in $S_{BH}$ due to Hawking radiation. In analogy with a classical blackbody, a cold large mass black hole in a warm thermal bath will absorb energy at a rate

$$\frac{dE}{dt} = \frac{\pi^2 k^4}{60\hbar^3 c^2}\sigma_S T_{R+M}^{\ 4} \qquad (19)$$

(and emit radiation $\frac{dE}{dt} = \frac{\pi^2 k^4}{60\hbar^3 c^2}\sigma_S T_{BH}^{\ 4}$) per polarization or helicity eigenstate where $\sigma_S = 27\pi G^2 M^2 / c^4$ is the geometrical optics cross-section[7]. Since the entropy of the background is maximized for a thermal bath, a thermal bath will give the strictest accretion constraint on $\Delta S$. For accretion, Eq (8) is now replaced by

$$\frac{dM}{dt} = +\frac{\beta_{R+M}\hbar c^4}{G^2 M_{R+M}^{\ 2}}\left(\frac{M}{M_{R+M}}\right)^2 \qquad (20)$$

where $\beta_{R+M} \sim 10^{-4}$ and $M_{R+M}$ is the mass of a black hole whose temperature equals the ambient temperature. Returning to Eq (7), the $de/dt$ term is now of order the first (absorption) term when

$$Q'_{1\approx 2} = \left\{\frac{\hbar c^4 \beta_{R+M}}{GM_{R+M}(e^{-1}de/dt)}\right\}^{1/2}\left(\frac{M}{M_{R+M}}\right)^{3/2}\left(2-\frac{\hbar c^4 \beta_{R+M}}{G^2 M_{R+M}M^2(e^{-1}de/dt)}\right)^{1/2} \qquad (21)$$

Provided $M \lesssim 5\times 10^{20}(M_{R+M}/M_{CMB})^4 M_\odot$, $Q_{MAX} \lesssim Q'_{1\approx 2}$ if $de/dt \approx 10^{-23}e$ per second. Gibbons[10] has argued that large black holes should not acquire significant charge. Once $Q/G^{1/2}M > G^{1/2}m_e/e \approx 5\times 10^{-22}$, a black hole can only gravitationally accrete a particle of like charge if the particle is projected at the black hole with an initial velocity[10]. A large black hole would be more likely to lose charge by accreting a particle of opposite charge. More rigorously, the Generalized Third Law of Thermodynamics states[9,16] that $T_{BH} = 0$, and hence $Q_{MAX}$, is not achievable by a finite sequence of steps. Thus the black hole charge will remain below $Q_{MAX}$ and the accretion from the radiation background, which depends on $T_{R+M}^4$, will always dominate the $de/dt$ term. Addressing $M \gtrsim 5 \times 10^{20} (M_{R+M} / M_{CMB})^4 M_\odot$, such supermassive black holes can not exist in the present Universe. Such a black hole would have a Schwarzschild radius of 25 – 50 Mpc, which is at least 1% of the current cosmic horizon, and be at least ten orders of magnitude more massive than an AGN core. Additionally, a black hole can only form when the size of the Universe is greater than the Schwarzschild radius of the hole and any black hole should produce noticeable distortion in the surrounding space-time out to at least about ten times its Schwarzschild radius. No such distortion on such a scale is observed in the present Universe. Stated another way, the existence of such supermassive black holes is ruled out by the present age and structure of the Universe. However even if such a supermassive black hole did exist it would presumably also discharge quickly by accretion of charge from its environment. Thus $Q'_{l=2}$ is not attainable by any black hole in the known Universe and even in the ultra-massive limit $\Delta S \geq 0$ is not violated.

In the special case when $T_{R+M} \geq T_{BH} \geq T_{CMB}$, $\Delta S$ due to absorption again must be greater than $\Delta S$ due to emission and, as we have shown in Case (I), $\Delta S$ due to emission is greater than or equal to the decrease in entropy due to $de/dt \approx 10^{-23} e$ per second for $T_{BH} \geq T_{CMB}$.

Hence for all $T_{BH} \leq T_{R+M}$, an increase in $e$ of the size indicated by the Webb et al data must produce a net increase in the generalized entropy, i.e. $\Delta S_{BH} + \Delta S_{R+M} \geq 0$ for all black holes with a net absorption.

\*          \*          \*

Combining Cases (IA), (IB) and (II), we conclude that if the electronic charge $e$ increases at a rate consistent with the Webb et al observations, the Generalized Second Law of Thermodynamics is not violated by Hawking black holes in the present Universe. In fact our analysis shows that a change in the fine structure constant of $\Delta \alpha / \alpha \sim 2 \times 10^{-23}$ per second corresponds to the maximum increase in $e$ allowed by the Generalized Second Law of Thermodynamics for black holes in the present Universe. The Second Law would be violated by emitting black holes if the Universe were only somewhat colder than today.

Extending our analysis to rotating black holes is straightforward and does not modify our conclusions. For a rotating, charged (Kerr-Newman) black hole, the $M + \sqrt{M^2 - Q^2/G}$ factor is replaced by $M + \sqrt{M^2 - Q^2/G - c^2 J^2/G^2 M^2}$. The maximal rotation is $J_{MAX} = GM^2 \sqrt{1 - Q^2/GM^2}/c$ and $J_{MAX} \approx GM^2/c$ unless $Q$ is very close to $Q_{MAX} = G^{1/2} M$ in which case the above treatment of $Q_{MAX}$ should be followed. Page has shown[9] that from $J = 0$ to $J = GM^2/c$, the power of a black hole emitting 4 spin-1/2 (3 neutrino species and electrons), 1 spin-1 (photon) and 1 spin-2 (graviton) species increases by a factor of 300 and the black hole loses spin faster than it loses energy. In the case of a $J \approx GM^2/c$ or $J = GM^2/c$ black hole, the superradiant mechanism for bosonic and fermionic modes dominates[9,13,14]. By the superradiant mechanism, which we discussed above for charged fermionic modes, a particle-antiparticle pair is created in the ergosphere with one particle with positive energy escaping to infinity and the other particle with locally positive energy being absorbed by the black hole. This mechanism has the consequence of increasing both the entropy of the environment and the entropy of the black hole, and spinning down the extremal black hole. In the case of a non-extremal black hole, the Generalized Third Law of Thermodynamics can also be applied to show that an existing black hole can not be spun up to $J_{MAX}$. Therefore the strictest constraints we obtain by including the $de/dt$ term in the Generalized Second Law of Thermodynamics come from black holes with $J = 0$ and charge $Q > 0$. Our conclusions are also not affected by the changes due to $de/dt$ in the Hawking and pair production discharge rates which are second order effects.

It should be noted that our derivation is essentially Standard Model physics and does not invoke quantum gravity. The black hole entropy and temperature, as defined, are required for *classical* General Relativity to be consistent with *classical* Thermodynamics[4,17,18]. Additionally the superradiant mechanism was first described[13,14] by Zel'dovich for classical black holes prior to the discovery of Hawking radiation. Schwinger pair production[19] is a non-perturbative process in standard QED.

If the Webb et al. measurements are correct, our analysis suggests at least two possibilities. We postulate that nature is such that $e$ varies at the maximal rate allowed by the Generalized Second Law of Thermodynamics. If this is so then, as seen in Eqs (10) and (12), the rate of increase in $\alpha$ should weaken with time as the Universe cools and $M_{CMB}$ increases. This postulate could be expanded if the increase in $\alpha$ is due not solely to $e$ varying, but to $e$, $\hbar$ and/or $c$ varying dependently as proposed in some Standard Model extensions, to say that the combined variation occurs at the maximal rate allowed by the Generalized Second Law of Thermodynamics. The maximal variation postulate should be explored theoretically and experimentally.

Alternatively perhaps the increase in $\alpha$ is due to a previously-undescribed coupling between the electron charge and the photon background whose effect is to partially screen the bare electron charge. As the Universe cools, the coupling weakens, increasing $e$. Although we have derived our result by applying the Generalized Second Law of Thermodynamics to black holes, in doing so we may have mathematically mimicked the relevant cosmological calculation: since the photon background is cosmological in origin it implicitly contains $G$. It should be investigated whether such a coupling arises as a higher order effect in standard QED or Standard Model extensions.

Acknowledgements
It is a pleasure to thank the University of Cambridge for hospitality.